\documentclass{vgtc}                          

\graphicspath{{figures/}{pictures/}{images/}{./}} 
\usepackage{amsfonts}

\newcommand{\Rspace}        {{\mathbb R}}
\usepackage{times}                     

\usepackage{tabu}                      
\usepackage{booktabs}                  
\usepackage{lipsum}                    
\usepackage{mwe}                       

\usepackage{mathptmx}                  

\usepackage{subcaption}
\usepackage{amsmath}

\newcommand\blfootnote[1]{%
  \begingroup
  \renewcommand\thefootnote{}\footnote{#1}%
  \addtocounter{footnote}{-1}%
  \endgroup
}
\onlineid{1024}

\vgtccategory{Research}

\vgtcinsertpkg

\title{Efficient Probabilistic Visualization of Local Divergence of 2D Vector Fields with Independent Gaussian Uncertainty}

\author{Timbwaoga A. J. Ouermi\thanks{e-mail: touermi@sci.utah.edu}\\ %
        \scriptsize SCI Institute, University of Utah %
\and Eric Li\\
     \scriptsize Indiana University %
\and Kenneth Moreland\\
     \scriptsize Oak Ridge National Laboratory %
\and Dave Pugmire\\
     \scriptsize Oak Ridge National Laboratory %
\and Chris R. Johnson\\
     \scriptsize SCI Institute, University of Utah %
\and Tushar M. Athawale\\
     \parbox{1.4in}{\scriptsize \centering Oak Ridge National Laboratory}}

\teaser{
\vspace{-4mm}
  \centering
  \includegraphics[width=0.9\linewidth]{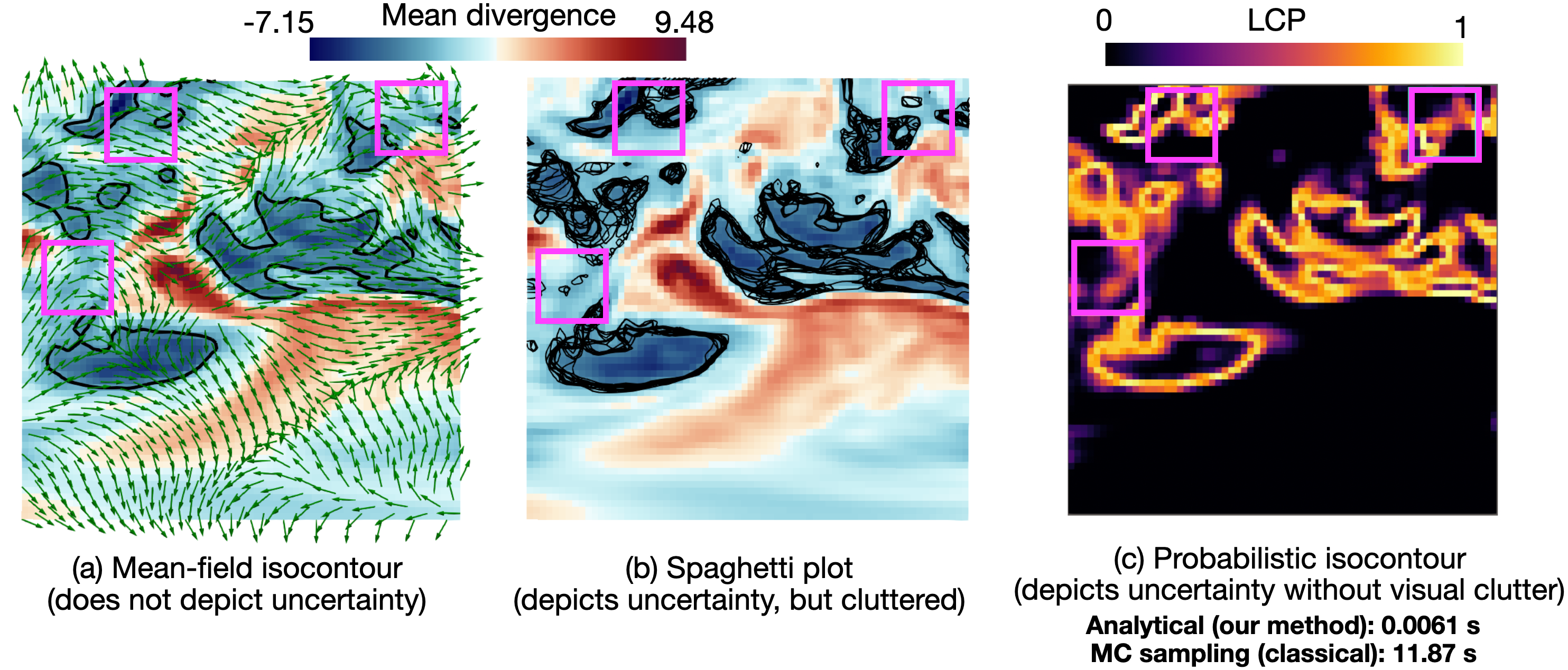}
  \vspace{-2mm}
  \caption {Mean-field (a) vs. probabilistic (b-c) isocontour visualization of the divergence of the wind vector field ensemble~\cite{Vitart2017} for the divergence isovalue $-2.525$. The mean divergence isocontour depicted in black (a) delineates potential positions of sink or negative divergence (the dark blue regions), but it does not depict its positional uncertainty, leading to potential errors in analysis. The spaghetti plot of isocontours (b) extracted from all ensemble members~\cite{TA:Potter:2009:ensemble-vis} reveals new isocontour positions that are vanished or truncated in the mean-field isocontour (illustrated by the magenta boxes). However, the spaghetti plot approach can lead to significant visual clutter. The LCP map~\cite{pothkowIndependentGaussianDataIsocontours, TA:Pothkow:2011:probMarchingCubes} (c) visualizes the probabilistic isocontour positions and clearly reveals positions of isocontour uncertainty (again, illustrated by magenta boxes). Our proposed analytical approach to derive the LCP map is highly efficient, providing a speed-up of over $1946\times$, compared to the classical MC sampling method.}

  \label{fig:teaser}
}

\abstract{
 This work focuses on visualizing uncertainty of local divergence of two-dimensional vector fields. Divergence is one of the fundamental attributes of fluid flows, as it can help domain scientists analyze potential positions of sources (positive divergence) and sinks (negative divergence) in the flow. However, uncertainty inherent in vector field data can lead to erroneous divergence computations, adversely impacting downstream analysis. While Monte Carlo (MC) sampling is a classical approach for estimating divergence uncertainty, it suffers from slow convergence and poor scalability with increasing data size and sample counts. Thus, we present a two-fold contribution that tackles the challenges of slow convergence and limited scalability of the MC approach. (1) We derive a closed-form approach for highly efficient and accurate uncertainty visualization of local divergence, assuming independently Gaussian-distributed vector uncertainties. (2) We further integrate our approach into Viskores, a platform-portable parallel library, to accelerate uncertainty visualization. In our results, we demonstrate significantly enhanced efficiency and accuracy of our serial analytical (speed-up up to $1946 \times$) and parallel Viskores (speed-up up to $19698 \times$) algorithms over the classical serial MC approach. We also demonstrate qualitative improvements of our probabilistic divergence visualizations over traditional mean-field visualization, which disregards uncertainty. We validate the accuracy and efficiency of our methods on wind forecast and ocean simulation datasets.
} 


\begin{document}


\firstsection{Introduction}

\maketitle

Divergence is one of the fundamental characteristics of the vector field data. Intuitively, it represents the amount of net flow in a local neighborhood. Visualization and analysis of divergence, therefore, can help domain scientists analyze the positions of sources (positive divergence) and sinks (negative divergence) within the vector field. \blfootnote{This manuscript has been authored by UT-Battelle, LLC under Contract No. DE-AC05-00OR22725 with the U.S. Department of Energy. The publisher, by accepting the article for publication, acknowledges that the U.S. Government retains a non-exclusive, paid up, irrevocable, world-wide license to publish or reproduce the published form of the manuscript, or allow others to do so, for U.S. Government purposes. The DOE will provide public access to these results in accordance with the DOE Public Access Plan (\url{http://energy.gov/downloads/doe-public-access-plan}).} Uncertainty in acquired vector field data, however, can propagate into the divergence computation, possibly leading to errors in downstream analysis. Uncertainty is inherent in simulation and measurement data stemming from a variety of factors, including fixed-bit precision, discretization errors, model assumptions, and parameter selection~\cite{kamal2021recent, TA:2004:Johnson:topScivis,TA:Potter:2012:UQtaxonomy}. Visual communication of uncertainty is important in mitigating data misrepresentation and errors in analysis. 
In this paper, we study the propagation of uncertainty in divergence computation for robust visualization and analysis of uncertain vector fields.

Monte Carlo (MC) is a well-established paradigm for quantifying and visualizing uncertainty~\cite{TA:Pothkow:2011:probMarchingCubes,Petz2012criticalPointProbability}. However, MC methods suffer from a major performance bottleneck: a large number of samples of the underlying probability distribution are required for convergence to a true solution. Thus, MC sampling methods are computationally expensive and do not scale well with increasing sample count and data size, rendering them ineffective in practice.

Multiple recent efforts have addressed the performance and scalability bottlenecks of MC solutions. In particular, three main directions are pursued. First, fast and accurate analytical solutions of uncertainty quantification for important data features (critical points~\cite{criticalPointUncertainty} and isosurfaces~\cite{pothkowIndependentGaussianDataIsocontours, TA:Athawale:2013:lerpUncertainty}) have been derived for probabilistic data models. Second, parallel algorithms and hardware acceleration techniques have been devised to improve computational performance for uncertainty visualization~\cite{Wang2023, funm2cHari}. Third, artificial neural networks have been investigated as surrogate models to accelerate uncertainty estimation~\cite{TA:2022:Han, surfaceBoxplot}.  Inspired by these efforts, we propose a novel analytical solution combined with parallelization for accurate, efficient, and scalable quantification and visualization of divergence uncertainty.

In summary, we analytically derive the divergence uncertainty when uncertainty in the vector field is characterized by independent Gaussian noise (a common noise model used in multiple prior studies~\cite{pothkowIndependentGaussianDataIsocontours,TA:Athawale:2021:topoMappingUncertaintyMarchingCubes}). We present a parallel algorithm using the Viskores library (previously known as vtk-m~\cite{TA:2016:Moreland:vtkm}) and utilize hardware acceleration (OpenMP and GPU) to speed-up computation. We showcase significantly improved accuracy and substantial speed-up/scalability of our serial and parallel analytical techniques over the serial MC approach. We validate and demonstrate the effectiveness of our methods on the wind and ocean simulation datasets.

\section{Related Work}\label{sec:related_work}
The field of uncertainty visualization has significantly progressed in the past two decades. Given the critical connection of uncertainty with the reliability of conclusions and decision-making, significant research has focused on studying uncertainty propagation in scalar field visualization algorithms. Examples include uncertain isosurfaces~\cite{TA:Pothkow:2011:probMarchingCubes, TA:Athawale:2013:lerpUncertainty, TA:Athawale:2016:nonparametricIsosurfaces}, uncertain direct volume rendering~\cite{TA:Shusen:2012:GMMdvr, TA:Athawale:2021:nonparametricDVR}, and uncertain topological visualizations
~\cite{TA:2013:Pfaffelmoser:gradientUncertainty,TA:Liebmann:2016:uncertainCriticalPoints,TA:Athawale2022MsComplex}.
However, research in uncertainty visualization of vector and tensor field data is still in the early stages, largely due to the added challenges posed by the higher dimensionality and complexity of these data~\cite{TA:Potter:2012:UQtaxonomy}.

We discuss a few works in vector field uncertainty visualization. A majority of existing methods have modeled uncertainty in vector field data with multivariate Gaussian distribution and developed MC sampling algorithms to convey positional uncertainty in important flow features, such as critical points~\cite{Petz2012criticalPointProbability}, vortex positions~\cite{otto2012vortex}, particle densities~\cite{otto2010uncertain2D,otto2010uncertain3D}, and finite-time Lyapunov exponent (FTLE)~\cite{TA:Guo:2016:LyapunovUncertaintyUnsteadyFlow}. Other approaches  for directly extracting flow features from ensemble data include a Helmholtz-Hodge decomposition~\cite{RIBEIRO201680}, Approximate Parallel Vectors (APV)~\cite{Tim20218}, and a generalization of boxplot \cite{TA:Whitaker:2013:contourBoxPlots,Mirzargar2014} to contours and curves.
Botchen et al.~\cite{textureFlowUncertainty} proposed a novel texture-based streamline smearing technique to convey uncertainty of vector fields. Several glyph-based approaches and novel glyph designs have been devised~\cite{ouermi2024glyphbaseduncertaintyvisualizationanalysis,Hlawatsch2011,pang1997approaches, TA:Lodha:1996:flowVisUncertainty} to convey uncertainty in magnitude and direction of vector data. In this work, we present a closed-form framework for visualization of local divergence uncertainty of vector fields for the independent Gaussian uncertainty model.
\section{Methods}
\label{sec:methods}

Let a discrete 2D vector field be represented as $\mathbf{f}:\Rspace^2 \rightarrow \Rspace^2$ with $\mathbf{f} = (u, v)$, where $u$ and $v$ denote horizontal ($x$) and vertical ($y$) components of vectors. The local divergence of a vector field then corresponds to a scalar quantity $\nabla \cdot \mathbf{f} = \frac{\partial u}{\partial x} +  \frac{\partial v}{\partial y}$. The positive divergence represents larger outgoing flow compared to incoming flow (and hence, a potential source). Conversely, the negative divergence represents a larger incoming flow compared to outgoing flow (and hence, potential sink). 


For vector field data with uncertainty, the data can be represented as a random variable $\mathbf{F} = (U, V)$. In this paper, we assume that the uncertain data are sampled on a uniform grid with grid spacing $\Delta x$ and $\Delta y$. Furthermore, we assume that each uncertain vector component (i.e., $U$ and $V$) has an independent Gaussian distribution $\mathcal{N}(\mu, \sigma^{2})$ with mean $\mu$ and standard deviation $\sigma$. Given the aforementioned assumptions, we derive the divergence uncertainty $\nabla \cdot \mathbf{F}$.

\subsection{Analytical Probabilistic Divergence} \label{sec:analytical_div_uncertainty}
Let $U_{i} \sim \mathcal{N}(\mu_{i}, \sigma_{i}^{2})$ and $V_{j} \sim \mathcal{N}(\mu_{j}, \sigma_{j}^{2})$ be the Gaussian model of the horizontal and vertical components of uncertain vector field at a pixel $(i,j)$. Similarly, let $U_{i-1}\sim \mathcal{N}(\mu_{i-1}, \sigma_{i-1}^{2})$, $U_{i+1}\sim \mathcal{N}(\mu_{i+1}, \sigma_{i+1}^{2})$, $V_{j-1}\sim \mathcal{N}(\mu_{j-1}, \sigma_{j-1}^{2})$, and $V_{j+1}\sim \mathcal{N}(\mu_{j+1}, \sigma_{j+1}^{2})$ denote vector component distributions at local neighbors. The divergence uncertainty at a pixel with 2D index $(i,j)$ is then approximated based on local neighbors according to:
\begin{equation}\label{eq:flow-div}
    \nabla \cdot \mathbf{F}_{i,j} = \frac{U_{i+1}-U_{i-1}}{2 \Delta x} + 
    \frac{V_{j+1}-V_{j-1}}{2 \Delta y}.
\end{equation} 
At the boundaries, the central finite differences in \cref{eq:flow-div} are replaced with forward and backward differences. Using the addition and multiplicative properties of Gaussian distributions, the resulting divergence is also a Gaussian distribution $\nabla \cdot \mathbf{F}_{i,j}\sim \mathcal{N}(\mu_{i,j}, \sigma^2_{i,j})$ where the mean and variance are expressed according to:
\begin{equation}\label{eq:div-mean}
    \mu_{i,j} = \frac{\mu_{i+1}}{2\Delta x} - \frac{\mu_{i-1}}{2\Delta x} + \frac{\mu_{j+1}}{2\Delta y}-\frac{\mu_{j-1}}{2\Delta y}, \textrm{ and}
\end{equation}
\begin{equation}\label{eq:div-variance}
    \sigma_{i,j}^{2} = \bigg(\frac{\sigma_{i+1}}{2\Delta x}\bigg)^{2} + \bigg(\frac{\sigma_{i-1}}{2\Delta x}\bigg)^{2} + \bigg( \frac{\sigma_{j+1}}{2\Delta y}\bigg)^{2} + \bigg(\frac{\sigma_{j-1}}{2\Delta y}\bigg)^{2}.
\end{equation}

The closed-form expressions in~\cref{eq:div-mean} and~\cref{eq:div-variance} provide an accurate and computationally efficient method compared to brute force MC sampling that can be used to estimate $\hat{\mu}_{i,j}$ and $\hat{\sigma}^2_{i,j}$.

Having derived $\nabla \cdot \mathbf{F}_{i,j} \sim \mathcal{N}(\mu_{i,j}, \sigma^2_{i,j})$ in closed-form at each grid vertex, we feed the derived Gaussian distribution to probabilistic isocontour visualization algorithm proposed by P\"{o}thkow et al.~\cite{pothkowIndependentGaussianDataIsocontours} for efficient visualization of level-crossing probability (LCP). Specifically, LCP quantifies the probability of isocontour to cross a grid cell given Gaussian-distributed data at cell vertices (in our case, $\mathcal{N}(\mu_{i,j}, \sigma^2_{i,j})$). We then visualize LCP via colormapping.

\subsection{Divergence Uncertainty Validation} \label{sec:div_uncertainty_validation}

\begin{figure}[t]
    \centering
    \includegraphics[width=0.45\linewidth]{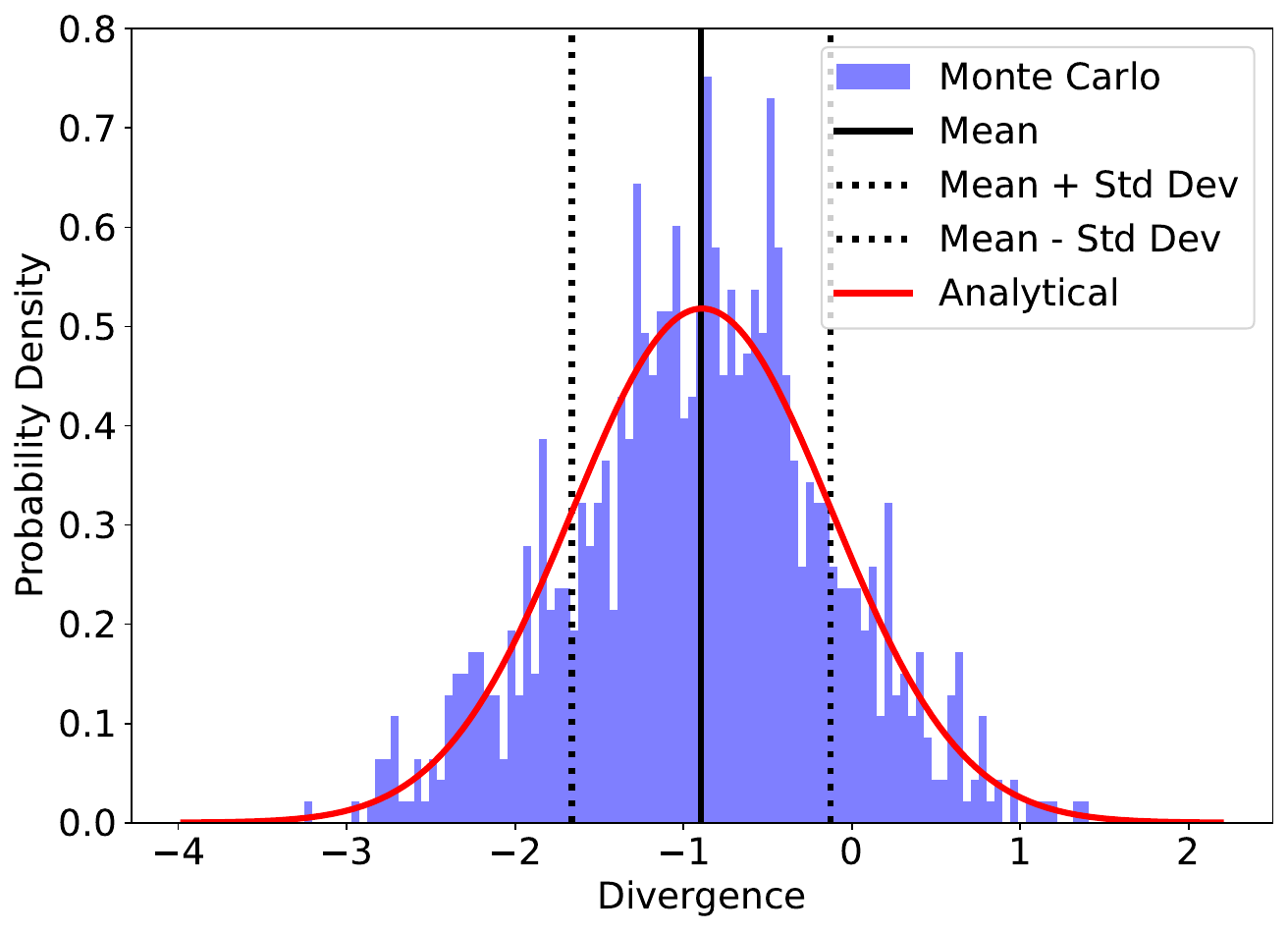}
    \includegraphics[width=0.45\linewidth]{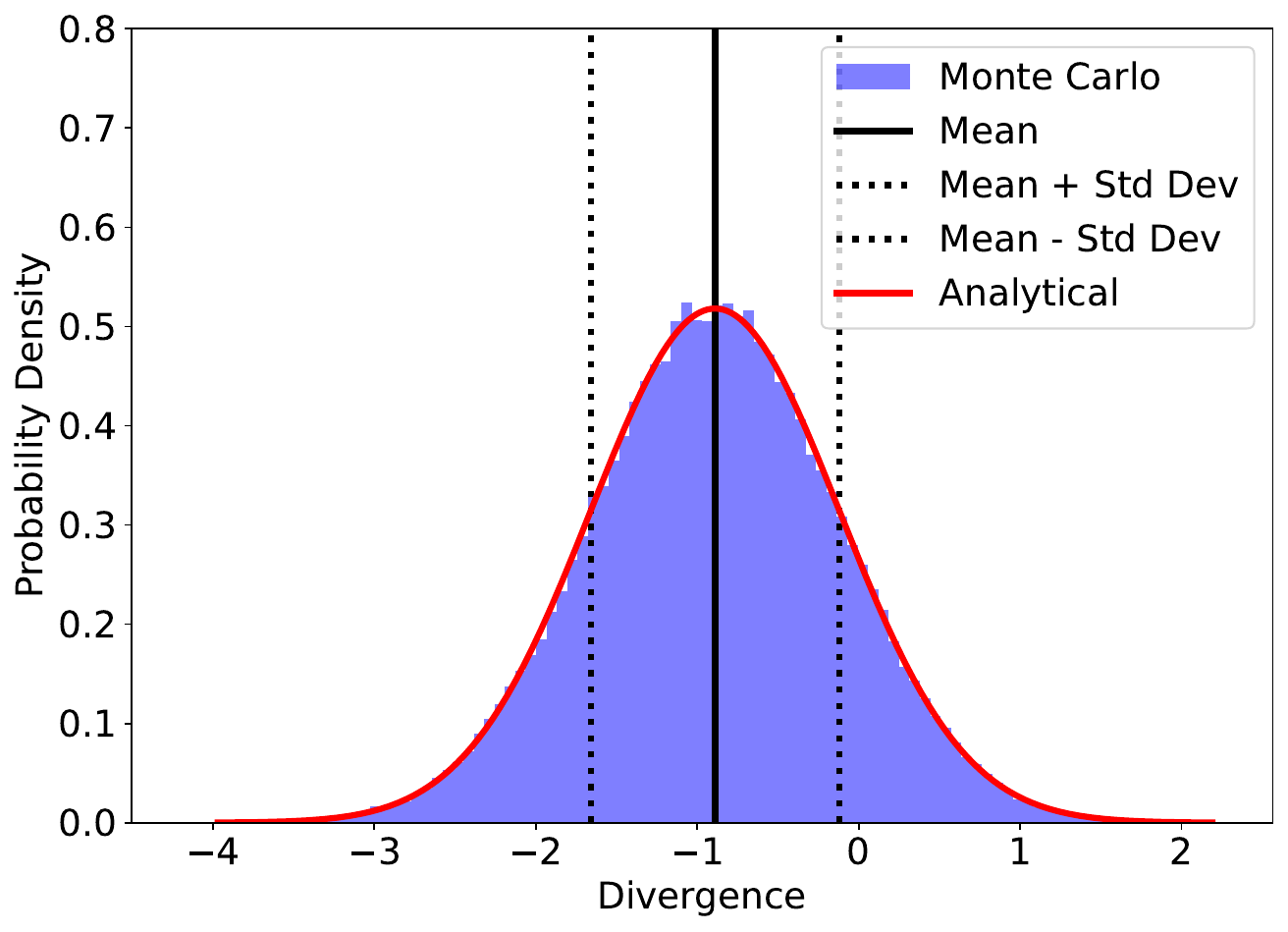}
    \begin{tabular}{c c c  c c}
            & $U_{i-1}$     &  $U_{i+1}$     & $V_{i-1}$     &  $V_{i+1}$    \\
            \hline
            $\mu$$(\sigma)$ &  $5.98$$(0.96)$   & $6.40$$(0.38)$      & $6.50$$(0.94)$      &   $4.30$$(0.65)$  \\
    \end{tabular}
     \begin{tabular}{c c c c}
         samples   & 1000    & & 100000    \\
         \hline
            $e_{m}(\sigma_{m})$   & $8.6e-3$$(2.3e-3)$  & &  $8.0e-4$ $(8.0e-4)$  \\
    \end{tabular}
    \vspace{-2mm}
    \caption{MC (blue curve) vs. analytical (red curve) divergence probability distribution. The mean and standard deviation values used are listed below the plots. The MC results are plotted for $1e+3$ (left column) and $1e+5$ (right column) samples.}
    \label{fig:divergence-validation}
\end{figure}

We validate our analytical derivation for divergence probability distribution (i.e.,~\cref{eq:div-mean} and~\cref{eq:div-variance}) against the MC method through a simple 1D experiment. Specifically, the parameters of distributions for uncertain vector components at local neighbors, i.e., $U_{i-1}$, $U_{i+1}$, $V_{j-1}$, and $V_{j+1}$, are selected randomly (see the parameters listed in~\cref{fig:divergence-validation}). Then the results of the divergence probability distribution at a pixel $(i,j)$ computed using the MC and proposed analytical method are compared for validation in~\cref{fig:divergence-validation}.

In~\cref{fig:divergence-validation}, the histograms are computed using the MC sampling approach with sample counts $1e+3$ (left column) and $1e+5$ (right column). In particular, for the MC sampling method, the vector field distribution at neighboring grid locations are sampled, and empirical divergence is calculated per sample according to~\cref{eq:flow-div}. The histogram is derived for the empirical divergence values. The red curves in~\cref{fig:divergence-validation} are plotted corresponding to analytical distribution $\mathcal{N}(\mu_{i,j}, \sigma^2_{i,j})$ (\cref{eq:div-mean} and~\cref{eq:div-variance}). The MC solution converges to the analytical solution as the sample count increases from $1e+3$ and $1e+5$. 
The quantitative difference between the empirical estimate and analytical solution for the mean ($e_{m}$) and the standard deviation ($e_{\sigma}$) is reported in the table at the bottom of~\cref{fig:divergence-validation}. The same shapes of the MC and analytical distributions and quantitative convergence validate our closed-form derivation.

\subsection{Viskores Parallel Algorithm\label{sec:viskores_parallel_algo}}
Given that the local probabilistic divergence computation is a per grid pixel (\cref{sec:analytical_div_uncertainty}), it is embarrassingly parallel. Thus, similar to the previous work by Wang et al.~\cite{Wang2023} and Hari et al.~\cite{funm2cHari}, we parallelize the divergence uncertainty computation with the Viskores library~\cite{TA:2016:Moreland:vtkm}.  Using our Viskores algorithm, we demonstrate significant speed-up in uncertainty visualization of local divergence.
\section{Results}\label{sec:results}

We demonstrate the performance, accuracy, and utility of our methods through experiments on wind forecast and ocean simulation datasets. We use isocontour visualization for analyzing divergence because it is a fundamental and widely used approach for exploration and analysis of scalar fields. Furthermore, isocountours of a divergence field help segment out sources and sinks in vector fields. In all experiments, the timing results are averaged over 10 independent runs for each experiment. 

\paragraph{\textbf{Wind dataset}~\cite{Vitart2017}:} The dataset is available from the IRI/LDEO Climate Data Library. The ensemble comprises 15 members representing the wind velocity, where all members are sampled on a grid of resolution $68 \times 68$.~\cref{fig:teaser} demonstrates results of our methods by visualizing uncertainty in the wind field's divergence to provide a robust representation of potential sink regions. 



~\cref{fig:teaser}a depicts the divergence of the mean of ensemble of velocity vector fields with an isocontour at the isovalue $-2.525$. Whereas the mean-field visualization provides a clean and deterministic view of the data, it completely obscures ensemble variability. In contrast, the spaghetti plot in \cref{fig:teaser}b, which overlays isocontours from all 15 members, effectively visualizes the spatial uncertainty and suggests likely presence of new features that are absent or truncated in the mean-field visualization, as illustrated by the magenta boxes. However, spaghetti plots suffer from significant over-plotting, thereby leading to a cluttered visualization. The probabilistic map in \cref{fig:teaser}c resolves this clutter by visualizing LCP~\cite{pothkowIndependentGaussianDataIsocontours}. The LCP visualization confirms the high likelihood of new sink features in the boxed region in a clear and quantifiable manner compared to the mean-field visualization that disregards uncertainty. 

The LCP visualization with classical MC sampling, however, suffers from a performance bottleneck based on the sample and grid size, a limitation that is addressed by our proposed analytical derivation (\cref{sec:analytical_div_uncertainty}). 
To demonstrate the computational advantage, we compare the performance of our method against the MC approach with 1,000 samples. The MC method takes $11.87$ s, whereas our proposed analytical solution took only $0.0061$ s (leading to a speed-up of up to $1946\times$) for serial implementation on the Apple M4 processor. Furthermore, our parallel algorithm (\cref{sec:viskores_parallel_algo}) with an OpenMP backend takes $0.0012$ s with $10$ processing threads, leading to a speed-up of over $9891\times$ compared to the serial MC algorithm. We elaborate on performance and accuracy curves for experiments on the Red Sea dataset given its larger size than the wind dataset.



\paragraph{\textbf{Red Sea dataset}~\cite{TA:2020:redSea}:}  The dataset is available on the 2020 IEEE SciVis contest website. The ensemble comprises $20$ members representing the velocity field, where all members are sampled on a grid of resolution $500 \times 500$. We demonstrate efficient uncertainty visualization using our methods to provide a more robust visual representation of Red Sea vortex positions. For oceanologists, knowledge of the positions of the vortices in the sea plays a critical role in understanding of energy and particle transport~\cite{TA:2020:redSea}. 

\begin{figure*} [t]
    \centering
    \includegraphics[width=0.9\linewidth]{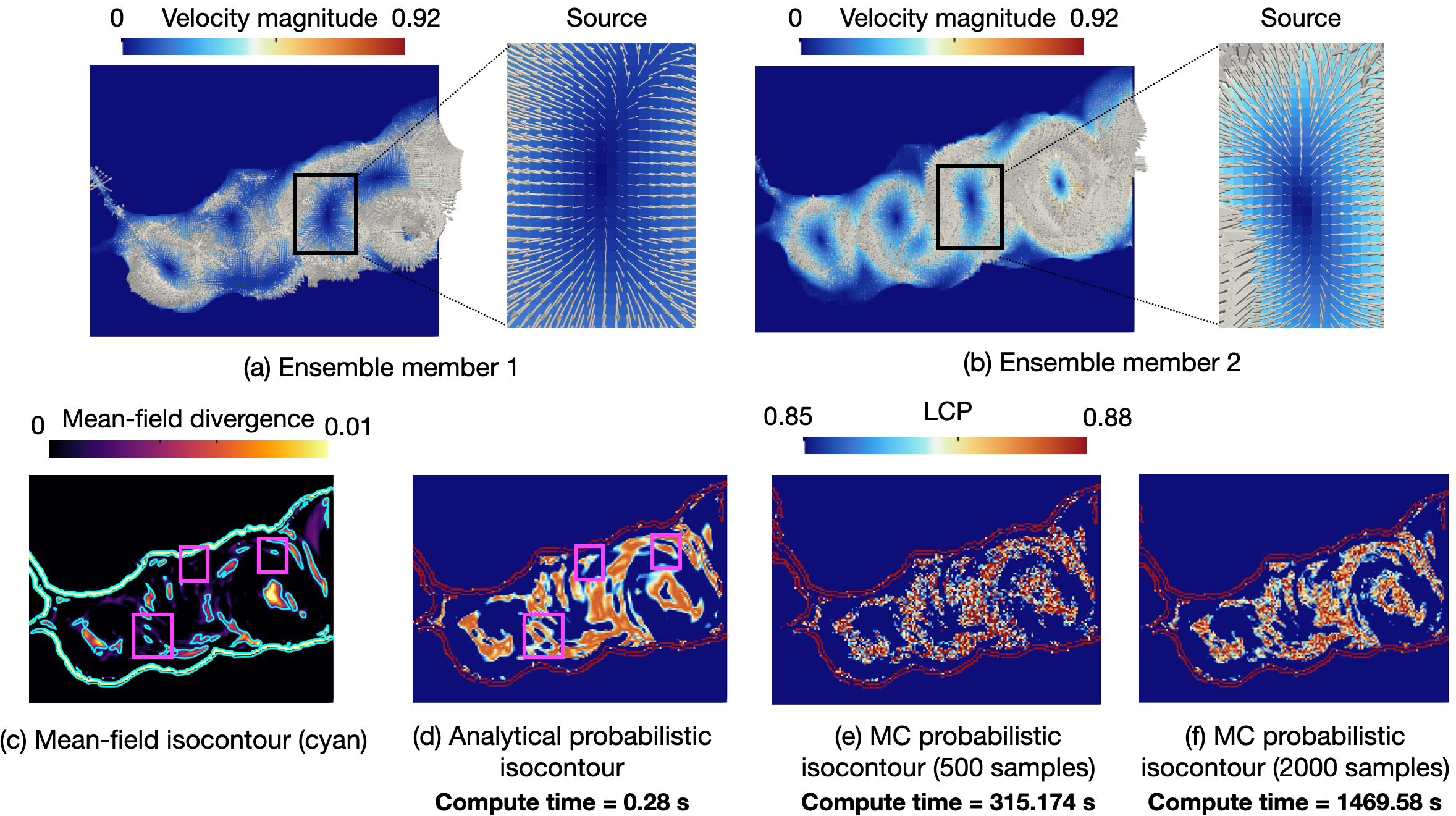}
    \vspace{-2mm}
    \caption{Visualization of divergence of the gradient of velocity magnitude field for the Red Sea dataset with 20 ensemble members. The top row (images a and b) depicts gradient fields for the two ensemble members visualized with the arrow glyphs. The zoomed-in views indicate source (positive divergence) positions in the gradient fields, which are representative of vortex core positions. However, the source positions vary considerably across the two ensemble members. Image c visualizes the divergence of the mean of ensemble of gradient fields with potential sources/vortex core positions (yellow/red regions) delineated using the isocontour (cyan) with the divergence isovalue $0.003$. The mean-field isocontour, however, does not account for uncertainty across the ensemble members. Image d visualizes probabilistic isocontour positions derived with our analytical approach. It reveals new potential positive divergence regions indicative of likely vortex cores that are missed or truncated in the mean-field visualization (as illustrated with the magenta boxes). The results in images e and f computed using the classical MC approach are much less accurate and slower than the analytical result in image d.}
    \label{fig:red-sea-data}
\end{figure*}

For our analysis, we compute the pixelwise gradient vectors of the velocity magnitude field for each ensemble member and analyze uncertainty of the divergence of the gradient fields to study the vortex positions. The results are shown in~\cref{fig:red-sea-data}. As seen in~\cref{fig:red-sea-data}a-b, the gradient fields for the two ensemble members clearly indicate the vortex positions with all gradient vectors pointing outward, representative of a source (positive local divergence).~\Cref{fig:red-sea-data}c visualizes colormapped local divergence of the mean of gradient fields with isocontour for the divergence isosvalue $0.003$ depicted in cyan. The isocontour segments out the regions of positive divergence (source) representative of vortex core. Note that the maximum divergence value in the mean-field is $0.14$ but large divergence values are only at a few pixels given sufficiently high resolution of the data (i.e., $500 \times 500$). Therefore, the maximum value of divergence is mapped to $0.01$ for colormapping in~\cref{fig:red-sea-data}c to prominently  highlight positive divergence (yellow/red) regions. 

\begin{figure}[!ht]
    \centering
    \includegraphics[width=0.46\linewidth]{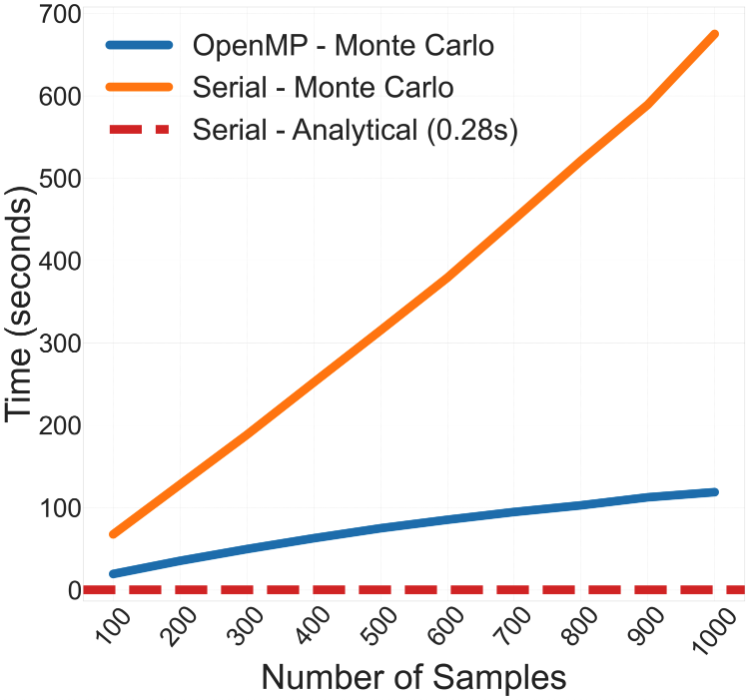}
    \includegraphics[width=0.46\linewidth]{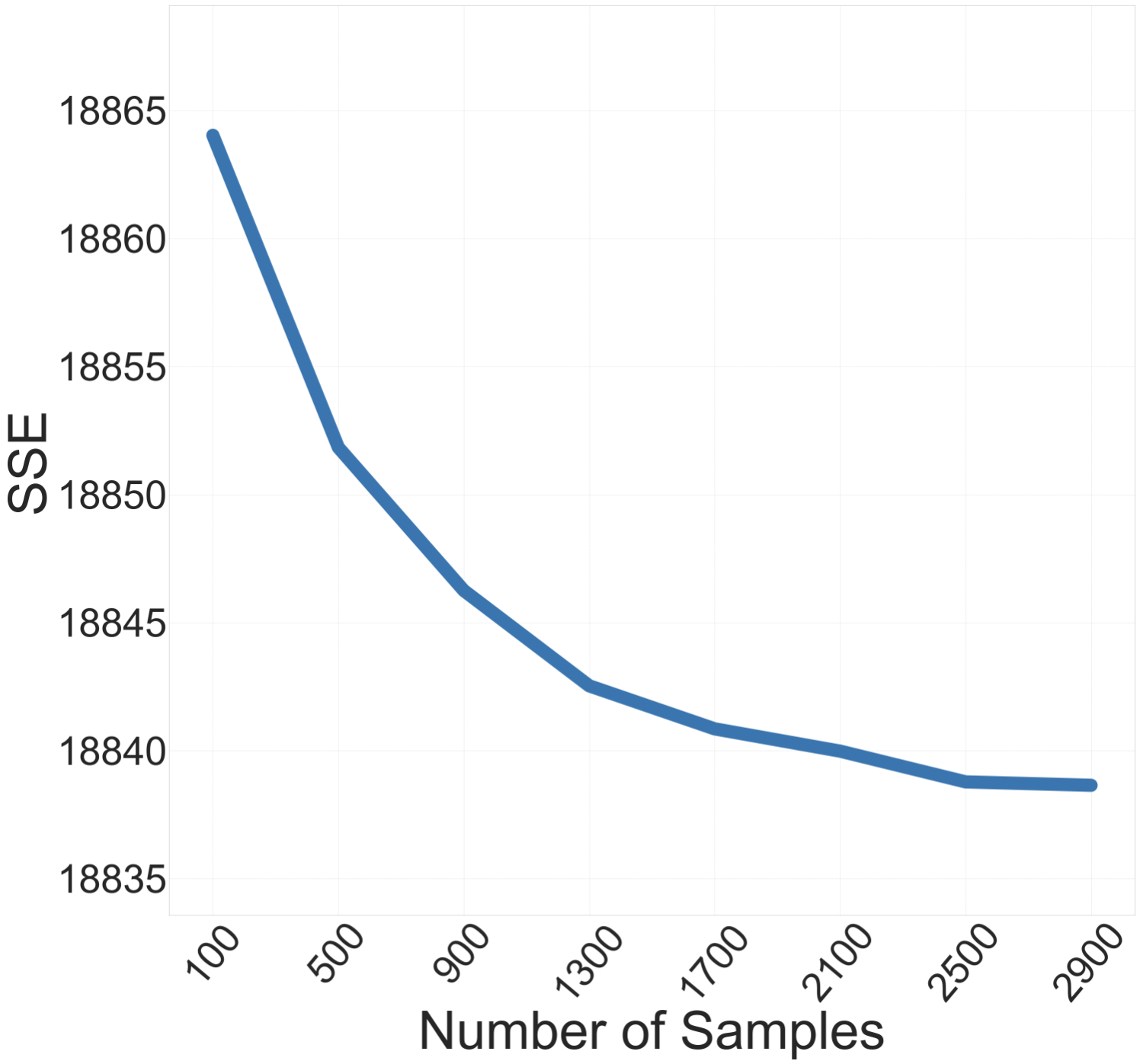}
    \vspace{-2mm}
    \caption{ Performance and accuracy plots for the Red Sea dataset. Our analytical method is far more efficient and accurate than the MC sampling method. }
    \label{fig:performance-comparison-red-sea}
\end{figure}

~\Cref{fig:red-sea-data}d-f visualize uncertainty with LCP. The analytically derived visualization in~\cref{fig:red-sea-data}d reveals the new potential positions of sources/vortex cores (illustrated by magenta boxes) that are vanished or truncated in the mean-field divergence visualization in~\cref{fig:red-sea-data}c. The visualizations in~\cref{fig:red-sea-data}e-f derived using $500$ and $2,000$ MC samples of uncertain vectors, respectively, are much slower to compute and are of lower quality than the analytically derived solution in~\cref{fig:red-sea-data}d. The result derived using $500$ samples took $315.174$s, whereas our proposed analytical solution took only $0.28$s (leading to a speed-up of $1125.62 \times$) for serial implementation on the Apple M4 processor. Furthermore, our parallel Viskores algorithm with OpenMP backend took $0.03$s with $10$ processing threads, leading to a speed-up of $10505.8 \times$ compared to the serial MC algorithm with 500 samples. We also ran our parallel Viskores algorithm on the AMD GPU on Oak Ridge National Laboratory's Frontier Supercomputer~\cite{Atchley2023}, which took $0.016$s and exhibited a speed-up of $19698.37 \times$ compared to the serial MC algorithm with $500$ samples. Only about $2 \times$ speed-up of AMD GPU over OpenMP is attributed to a relatively small data size to keep the GPU busy and the overhead related to the launch of the GPU kernel. For higher MC sample count, a further speed-up is observed.

~\Cref{fig:performance-comparison-red-sea} shows the plots of quantitative comparison of performance and accuracy for analytical and MC methods. Our method on a serial processor (the red dotted curve) is far more efficient and scalable compared to the MC sampling method (solid curves) with increasing sample count. Furthermore, the analytical method provides an accurate solution and the sum of squared errors (SSE) of the MC sampling solution with respect to the analytical solution  drops with an increase in the sample count. The performance and accuracy curves for the wind dataset (\cref{fig:teaser}) exhibit similar trends to those for the Red Sea dataset and are not explicitly shown.

\section{Conclusion and Future Work}
This paper addressed the critical challenge of quantifying and visualizing divergence uncertainty in vector fields, a fundamental characteristic to understand local net flow and identify sources and sinks. As standard MC sampling is computationally expensive for uncertainty estimations, we derive an analytical expression of the local flow divergence uncertainty (\cref{sec:analytical_div_uncertainty}) under the independent Gaussian assumption of vector data at each grid point. We parallelize our algorithms using the Viskores library (\cref{sec:viskores_parallel_algo}) and leverage hardware acceleration to further reduce the computational cost. We use our techniques on two simulation datasets to demonstrate the benefits of the divergence uncertainty visualization to identify potential sinks and sources that can be missed in deterministic mean-field visualization that disregards uncertainty in vector data. The proposed analytical method with parallelization achieved speedup up to $19698 \times$ over classical serial computation with MC Sampling, thus enabling efficient visualization of uncertainty. 

This work focuses on 2D vector fields with independent Gaussian noise assumptions and does not take into account correlation between vector components, spatial neighbors, and divergence. In future work, we plan to extend this work to take into account correlation and expand it to 3D vector fields. These improvements will broaden the scope of our approach, enabling more robust and generalizable uncertainty-aware divergence analysis for flow data.


\acknowledgments{
This work was partially supported by the Intel OneAPI CoE, the Intel Graphics and Visualization Institutes of XeLLENCE, and the DOE Ab-initio Visualization for Innovative Science (AIVIS) grant 2428225. This research used resources of the Oak Ridge Leadership Computing Facility at the Oak Ridge National Laboratory, which is supported by the Office of Science of the U.S. Department of Energy under Contract No. DE-AC05-00OR22725.}

\bibliographystyle{abbrv-doi}

\bibliography{template}
\end{document}